\documentclass[12pt]{article}
\usepackage[english,german,french,polish]{babel}
\usepackage[T1]{fontenc}

\begin{document}

\baselineskip 0.72cm
\topmargin -0.4in
\oddsidemargin -0.1in

\let\ni=\noindent

\renewcommand{\thefootnote}{\fnsymbol{footnote}}

\newcommand{\CKM}{Cabibbo-Kobayashi-Maskawa }

\newcommand{\Stk}{SuperKamiokande }

\newcommand{\SM}{Standard Model }

\pagestyle {plain}

\setcounter{page}{1}

\pagestyle{empty}




~~~~~
\begin{flushright}
IFT-02/23
\end{flushright}

\vspace{0.2cm}

{\large\centerline{\bf Explicit seesaw with nearly bimaximal}}

{\large\centerline{\bf neutrino mixing and no LSND effect{\footnote {Work supported in part by the Polish State Committee for Scientific Research (KBN), grant 5 P03B 119 20 (2001--2002).}}}}

\vspace{0.5cm}

{\centerline {\sc Wojciech Kr\'{o}likowski}}

\vspace{0.23cm}

{\centerline {\it Institute of Theoretical Physics, Warsaw University}}

{\centerline {\it Ho\.{z}a 69,~~PL--00--681 Warszawa, ~Poland}}

\vspace{0.5cm}

{\centerline{\bf Abstract}}

\vspace{0.3cm}

An explicit model of neutrino texture is presented, where in the $6\times 6$ mass matrix  the Majorana lefthanded component is zero, the Majorana righthanded component --- diagonal with equal entries, and the Dirac component gets a hierarchical structure, deformed by nearly bimaximal mixing. If the Majorana righthanded component dominates over the Dirac component, the familiar seesaw mechanism leads effectively to the popular, nearly bimaximal oscillations of active neutrinos. The Dirac component, before its deformation, may be similar in shape to the charged-lepton and quark mass matrices. Then, parameters for solar and atmospheric neutrinos may be related to each other, {\it predicting} from the SuperKamiokande value of $\Delta m_{32}^2$ a tiny $\Delta m_{21}^2$, typical for MSW LOW solar solution (rather than for MSW Large Mixing Angle solution). 

\vspace{0.4cm}

\ni PACS numbers: 12.15.Ff , 14.60.Pq , 12.15.Hh .

\vspace{0.8cm}

\ni July 2002

\vfill\eject

~~~~~
\pagestyle {plain}

\setcounter{page}{1}
 
\ni {\bf 1. Introduction.} As is well known, the popular nearly bimaximal form of mixing matrix for three active neutrinos $\nu_{e L}$, $\nu_{\mu L}$, $\nu_{\tau L}$ [1],

\begin{equation} 
U^{(3)} =   \left( \begin{array}{ccc} c_{12} & s_{12} & 0 \\ -s_{12}c_{23} & c_{12}c_{23} & s_{23} \\ s_{12}s_{23} & -c_{12}s_{23} & c_{23} \end{array} \right) \; ,
\end{equation} 

\ni arising from its generic shape {\it \`{a} la} \CKM [2] by putting $s_{13} = 0$ and $c_{12}\,,\, s_{12}\,,\, c_{23}\,,\, s_{23}$ not so far from $1/\sqrt{2}$, is globally consistent with neutrino oscillation experiments [3] for solar $\nu_e$'s and atmospheric $\nu_\mu$'s as well as with the negative Chooz experiment for reactor $\bar{\nu}_e$'s. It cannot explain, however, the possible LSND effect for accelerator $\bar{\nu}_\mu$'s that, if confirmed, may require the existence of one, at least, extra (sterile) light neutrino $\nu_{s L}$ (different, in general, from the conventional sterile neutrinos $(\nu_{e R})^c\,,\, (\nu_{\mu R})^c\,,\, (\nu_{\tau R})^c$). This sterile neutrino may appear in the so-called 2+2 or 3+1 version [3].

If active neutrinos $\nu_{\alpha L}\;(\alpha = e, \mu, \tau)$ are of Majorana type, their effective mass term in the Lagrangian has the form

\begin{equation} 
- {\cal L}^{(3)}_{\rm mass} = \frac{1}{2}\sum_{\alpha \beta} \overline{(\nu_{\alpha L})^c} M^{(3)}_{\alpha \beta} \nu_{\beta L} + {\rm h.\,c.} \;,
\end{equation} 

\ni where the (Majorana) mass matrix $M^{(3)} = \left( M^{(3)}_{\alpha \beta} \right)$ is symmetric due to the identity $\overline{\nu_{\alpha L}} (\nu_{\beta L})^c = \overline{\nu_{\beta L}} (\nu_{\alpha L})^c $ (here, the normal ordering of bilinear neutrino terms is implicit). In the flavor representation, where the charged-lepton $3\times 3$ mass matrix is diagonal, the generic neutrino $3\times 3$ mixing matrix $U^{(3)} = \left(U^{(3)}_{\alpha i}\right)$ is, at the same time, the unitary diagonalizing matrix for the neutrino $3\times 3$ mass matrix,

\begin{equation} 
U^{(3)\,\dagger} M^{(3)} U^{(3)} = {\rm diag}(m_1 \,,\,m_2 \,,\,m_3)
\end{equation} 

\ni with $m_1 \,,\,m_2 \,,\,m_3$ denoting neutrino masses (real numbers). This is true, {\it if} $M^{(3)}$ is not only symmetric but also real, {\it i.e.}, its possible two Majorana phases and one Dirac phase $\delta $ are trivial (and so, the possible CP violation is ignored for neutrinos). Then, $ M^{(3)}_{\alpha \beta}  = \sum_i U^{(3)}_{\alpha i} m_i U^{(3)\,*}_{\beta i}$ where $ U^{(3)}$ is orthogonal and real. In particular, for $ U^{(3)}$ given in Eq. (1) the Dirac phase $\delta $ is absent, due to $s_{13} = 0$ ({\it e.g.} $ U_{e 3} = s_{13} e^{-i\delta} = 0$). The active-neutrino flavor and mass fields, $\nu_{\alpha L}\;(\alpha = e, \mu, \tau)$ and $\nu_{i L}\;(i = 1,2,3)$, are related through the unitary transformation 

\begin{equation} 
\nu_{\alpha L}  = \sum_i U^{(3)}_{\alpha i}  \nu_{i L} \;,
\end{equation} 

\ni even if three CP violating phases are nontrivial. Note that CP violation in the neutrino oscillations may be caused only by the Dirac phase $\delta $ (if it is present in the mixing matrix $U^{(3)}$).

According to the popular viewpoint, the active-neutrino effective mass term (2) arises through the familiar seesaw mechanism [4] from the generic neutrino mass term

\begin{equation} 
- {\cal L}_{\rm mass} = \frac{1}{2} \sum_{\alpha \beta} \left( \overline{(\nu_{\alpha L})^c} \,,\, \overline{\nu_{\alpha R}}\right) \left( \begin{array}{cc} M^{(L)}_{\alpha \beta} & M^{(D)}_{\alpha \beta} \\ M^{(D)}_{\beta \alpha} & M^{(R)}_{\alpha \beta} \end{array} \right) \left( \begin{array}{c} \nu_{\beta L} \\ (\nu_{\beta R})^c \end{array} \right) + {\rm h.\,c.} 
\end{equation} 

\ni including both the active neutrinos $\nu_{\alpha L}$ and $(\nu_{\alpha L})^c$ as well as the (conventional) sterile neutrinos $\nu_{\alpha R}$ and $(\nu_{\alpha R})^c \;(\alpha = e\,,\,\mu\,,\,\tau)$. In the seesaw case, the Majo\-rana righthanded mass matrix $ M^{(R)} = \left(M^{(R)}_{\alpha \beta}\right)$ is presumed to dominate over the Dirac  mass matrix $ M^{(D)} = \left(M^{(D)}_{\alpha \beta}\right)$ that in turn  dominates over the Majorana lefthanded mass matrix $ M^{(L)} = \left(M^{(L)}_{\alpha \beta}\right)$ which is expected to be zero ($M^{(D)}$ and $M^{(L)}$, in contrast to    $M^{(R)}$, violate the electroweak symmetry SU(2)$\times$U(1); of the two, only the first may arise from the conventional doublet Higgs mechanism in a renormalizable way). Such a seesaw mechanism leads effectively to the active-neutrino (Majorana) mass matrix $ M^{(3)}$ appearing in the mass term (2). Then, $ M^{(3)} \simeq - M^{(D)}  M^{(R)\,-1}M^{(D)\,T}$, and so, $M^{(3)}$ is guaranteed to be small, while the (conventional) sterile neutrinos get approximately $M^{(R)}$ as their effective (Majorana) mass matrix and, therefore, are practically decoupled from the active neutrinos. Opposite to the seesaw case is the pseudo-Dirac case, when $M^{(D)}$ is presumed to dominate  over $M^{(R)}$ (and over the vanishing $M^{(L)}$)  [5]. Then,  $-M^{(D)}$ and $+M^{(D)}$  (or {\it vice versa}) become approximately the effective (Majorana) mass matrices for active and (conventional) sterile neutrinos, respectively. This implies $m_1 \simeq - m_4$, $m_2 \simeq - m_5$, $m_3 \simeq - m_6$ for the pseudo-Dirac neutrino mass spectrum.

In the present note, we study an explicit model for the overall $6\times 6$ mass matrix
 
\begin{equation}
M = \left( \begin{array}{cc} 0 & M^{(D)} \\ M^{(D)\,T}  & M^{(R)} \end{array} \right) 
\end{equation}

\ni appearing in the generic neutrino mass term (5). If in this model $M^{(R)}$ dominates over $M^{(D)}$, the familiar seesaw mechanism leads effectively to the popular, nearly bimaximal oscillations of active neutrinos. But, in this model, these nearly bimaximal oscillations hold also in the pseudo-Dirac case, when $M^{(R)}$ is dominated by $M^{(D)}$. 

\vspace{0.2cm}

\ni {\bf 2. Model.} Let us assume in Eq. (6) that  

\begin{equation} 
M^{(D)} = {\stackrel{0}{m}}\; U^{(3)}\frac{1}{2} \left( \begin{array}{ccc} \tan 2\theta_{14} & 0 & 0 \\ 0 & \tan 2\theta_{25} & 0 \\ 0 & 0 & \tan 2\theta_{36} \end{array} \right)
\end{equation} 

\ni and

\begin{equation} 
M^{(R)} = {\stackrel{0}{m}} \left(\begin{array}{ccc} 1 & 0 & 0 \\ 0 & 1 & 0 \\ 0 & 0 & 1 \end{array}\right) \;,
\end{equation}

\ni where ${\stackrel{0}{m}}\, > 0$ is a mass scale and

\begin{equation} 
\frac{1}{2}\tan 2\theta_{ij} = \frac{c_{ij}s_{ij}}{c^2_{ij} - s^2_{ij}} = \frac{t_{ij}}{1 - t^2_{ij}} \;(ij = 14,25,36)
\end{equation}

\ni denote three dimensionless parameters, connected with $c_{ij} = \cos \theta_{ij}$ and $s_{ij} = \sin \theta_{ij}$ or $t_{ij} = \tan \theta_{ij}$, while $U^{(3)}$ stands for the previous $3\times 3$ mixing matrix given in Eq. (1). Thus, the Dirac component $M^{(D)}$ of the overall neutrino mass matrix $M$ is a diagonal, potentially hierarchical structure, deformed by the popular, nearly bimaximal mixing matrix $U^{(3)}$ [6]. Evidently, in this $6\times 6$ model $ M^T = M$ and $M^* = M$ (the possible CP violation is ignored).

We claim that the unitary diagonalizing matrix $U$ for the overall $6\times 6$ mass matrix $M$ defined in Eqs. (7) and (8),

\begin{equation} 
U^\dagger M U = {\rm diag}(m_1\,,\,m_2\,,\,m_3\,,\,m_4\,,\,m_5\,,\,m_6)\;,
\end{equation} 

\ni gets the form

\begin{equation}
U = {\stackrel{1}{U}}{\stackrel{0}{U}} \;,\;{\stackrel{1}{U}} =  \left( \begin{array}{cc} U^{(3)} & 0^{(3)} \\ 0^{(3)}  & 1^{(3)} \end{array} \right) \;,\; {\stackrel{0}{U}} = \left( \begin{array}{cc} C^{(3)} & S^{(3)} \\ -S^{(3)} & C^{(3)} \end{array} \right)
\end{equation} 

\ni with $ U^{(3)}$ as given in Eq. (1) and

\begin{equation} 
1^{(3)} =  \left( \begin{array}{ccc} 1 & 0 & 0 \\ 0 & 1 & 0 \\ 0 & 0 & 1 \end{array} \right) \;,\; C^{(3)} =  \left( \begin{array}{ccc} c_{14} & 0 & 0 \\ 0 & c_{25} & 0 \\ 0 & 0 & c_{36} \end{array} \right)\;\; ,\;\;  S^{(3)} =  \left( \begin{array}{ccc} s_{14} & 0 & 0 \\ 0 & s_{25} & 0 \\ 0 & 0 & s_{36} \end{array} \right) \;.
\end{equation} 

\ni Evidently, $U^T = U^{-1}$ and $U^* = U$. Further, we claim that the neutrino mass spectrum takes the form

\begin{equation}
m_i = -{\stackrel{0}{m}}\frac{t^2_{ij}}{1 - t^2_{ij}} \;\;,\;\;m_j = {\stackrel{0}{m}} + {\stackrel{0}{m}} \frac{t^2_{ij}}{1 - t^2_{ij}} = \frac{\stackrel{0}{m}}{1 - t^2_{ij}} \;\;(ij = 14,25,36)\,.
\end{equation}

\ni Thus, $m_i + m_j = {\stackrel{0}{m}} $ and $m_i/m_j = - t^2_{ij}$.

The easiest way to prove the statement expressed by Eqs. (11) and (13) is to start with the diagonalizing matrix $U$ defined in Eqs. (11), (1) and (12), and then to show by applying the formula

\begin{equation} 
M =  U \, {\rm diag}(m_1\,,\,m_2\,,\,m_3\,,\,m_4\,,\,m_5\,,\,m_6)\, U^\dagger
\end{equation} 

\ni that the mass matrix $M$ is given as in Eqs. (6), (7) and (8), {\it if} the mass spectrum $m_1 ,\,m_2 ,\,m_3,\, m_4,\, m_5,\, m_6$ is taken in the form (13).

In the flavor representation, where charged-lepton mass matrix is diagonal, the $6\times 6$ diagonalizing matrix $U$ is, at the same time, the $6\times 6$ unitary mixing matrix relating three active and three (conventional) sterile flavor neutrino fields  $\nu_{\alpha L}\;(\alpha = e, \mu, \tau, e_s, \mu_s, \tau_s)$ with six mass neutrino fields $\nu_{i L}\;(i = 1,2,3,4,5,6)$: $\nu_{\alpha L} = \sum_i U_{\alpha i} \nu_{i L}$, where $\nu_{\alpha_s L} \equiv (\nu_{\alpha R})^c \;(\alpha = e, \mu, \tau)$.

It may be interesting to observe that the $6\times 6$ mass matrix $M$ defined in Eqs. (6), (7) and (8) can be presented as the unitary transform $ M = {\stackrel{1}{U}}{\stackrel{0}{M}} \stackrel{1}{U} \!^\dagger $ of the new simpler $6\times 6$ mass matrix
 
\begin{eqnarray} 
\stackrel{0}{M} = \left( \begin{array}{cc} 0 & \stackrel{0}{M}\!^{(D)} \\ \stackrel{0}{M}\!^{(D)\,T}  & \stackrel{0}{M}\!^{(R)} \end{array} \right) & ,& \nonumber \\
\;\stackrel{0}{M}\!^{(D)} = \stackrel{0}{m} \frac{1}{2}{\rm diag}(\tan 2\theta_{14},\tan 2\theta_{25},\tan 2\theta_{36}) & , & \stackrel{0}{M}\!^{(R)} = \,\stackrel{0}{m}\,1^{(3)} 
\end{eqnarray} 

\ni by means of $\stackrel{1}{U} =$ diag$\left( U^{(3)}, 1^{(3)} \right)$ [see Eqs. (7), (8) and (11)]. Thus, the Dirac component $\stackrel{0}{M}\!^{(D)}$ of $\stackrel{0}{M}$ (subject to the deformation by nearly bimaximal mixing) is potentially hierarchical. Before its deformation, this Dirac component $\stackrel{0}{M}\!^{(D)}$ may display a structure similar to the charged-lepton and quark $3\times 3$ mass matrices which, of course, are also of Dirac type.

In the seesaw mechanism [4] there appears an effective $6\times 6$ mass matrix $M_{\rm eff}$ approximately equal to the familiar block-diagonal form,

\begin{eqnarray} 
M_{\rm eff} & \simeq &  \left( \begin{array}{cc}  - M^{(D)} M^{(R)\,-1} M^{(D)\,T} & 0 \\ 0  & M^{(R)} \end{array} \right) \nonumber \\
 & = & \stackrel{1}{U} \left( \begin{array}{cc}  - \stackrel{0}{M}\!^{(D)} \stackrel{0}{M}\!^{(R)\,-1} \stackrel{0}{M} \!^{(D)\,T} & 0 \\ 0  & \stackrel{0}{M}\!^{(R)} \end{array} \right)  \stackrel{1}{U} \!^\dagger \;, 
\end{eqnarray} 

\ni where in our model $M^{(D)} = {U}^{(3)} \stackrel{0}{M}\!^{(D)}$, $\stackrel{0}{M}\!^{(D)}$ is given as in Eq. (15) and $M^{(R)}   = \stackrel{0}{M}\!^{(R)}  = {\stackrel{0}{m}} {1}^{(3)}$. Thus, from Eqs. (15) and (13)

\begin{eqnarray} 
 - \stackrel{0}{M}\!^{(D)} \stackrel{0}{M}\!^{(R)\,-1} \stackrel{0}{M}\!^{(D)\,T} & = & - \stackrel{0}{m} \frac{1}{2}{\rm diag}(\tan^2  2\theta_{14},\tan^2 2\theta_{25},\tan^2  2\theta_{36}) \nonumber \\
 & \simeq & -\stackrel{0}{m} {\rm diag}(t^2_{14},t^2_{25},t^2_{36}) \simeq {\rm diag}(m_1, m_2, m_3) \;, 
\end{eqnarray} 

\ni and

\begin{equation}
\stackrel{0}{M}\!^{(R)} = {\stackrel{0}{m}} 1^{(3)} \simeq {\rm diag}(m_4, m_5, m_6) \;,
\end{equation}

\ni since $t^2_{ij} \ll 1\;\;(ij = 14, 25, 36)$, what is the seesaw requirement (see Eqs. (13) giving $m_i/m_j = - t^2_{ij}$ and $m_j \simeq {\stackrel{0}{m}}$, the latter for $t^2_{ij} \ll 1$). From Eqs. (16), (17) and (18) we infer that

\begin{equation} 
M_{\rm eff} \simeq \stackrel{1}{U}{\rm diag} (m_1\,,\,m_2\,,\,m_3\,,\,m_4\,,\,m_5\,,\,m_6) \stackrel{1}{U} \!^\dagger \;.
\end{equation} 

\ni Comparing Eq. (19) with the formula (10), where $ U = \stackrel{1}{U} \stackrel{0}{U}$, and presenting $M$ as a unitary transform of $M_{\rm eff}$, $M = U_{\rm eff} M_{\rm eff} U^\dagger_{\rm eff}$, we obtain $U_{\rm eff} {\stackrel{1}{U}} \simeq \stackrel{1}{U} \stackrel{0}{U} $ and hence, the remarkable relation

\begin{equation} 
U_{\rm eff} \simeq \,\stackrel{1}{U} \stackrel{0}{U} \stackrel{1}{U} \!^\dagger = U \stackrel{1}{U} \!^\dagger  
\end{equation} 

\ni valid under the seesaw requirement ($t^2_{ij}\ll 1$). 

For the active-neutrino $3\times 3$ mass matrix appearing in the effective mass term (2) we get $M^{(3)} = M^{(L)}_{\rm eff} \simeq - M^{(D)} M^{(R)\,-1} M^{(D)\,T}$, if the seesaw mechanism works. As follows from Eqs. (16), (11) and (17), it is approximately diagonalised by means of the nearly bimaximal mixing matrix $U^{(3)}$ given in Eq. (1),

\begin{equation}
U^{(3)\,\dagger} M^{(L)}_{\rm eff} U^{(3)} \simeq - \stackrel{0}{M}\!^{(D)} \stackrel{0}{M}\!^{(R)\,-1} \stackrel{0}{M}\!^{(D)\,T} \simeq {\rm diag}(m_1, m_2, m_3) \;,
\end{equation}

\ni where $m_i \simeq - {\stackrel{0}{m}}\, t^2_{ij}\;\; (ij = 14, 25, 36)$. Thus, in the seesaw approximation the mixing matrix $U^{(3)}$ leads (in the vacuum) to the familiar, nearly bimaximal oscillation probabilities

\begin{eqnarray} 
P(\nu_e \rightarrow \nu_e)_{\rm sol}\;\;\;\, & = & 1 - (2c_{12}s_{12})^2 \sin^2 (x _{21})_{\rm sol}\;, \nonumber \\
P(\nu_\mu \rightarrow \nu_\mu)_{\rm atm} \;\;& = & 1 - (2c_{23}s_{23})^2 \left[ s^2_{12}\sin^2 (x _{31})_{\rm atm} + c^2_{12}\sin^2 (x _{32})_{\rm atm}\right] \nonumber \\
 & \simeq & 1 - (2c_{23}s_{23})^2 \sin^2 (x _{32})_{\rm atm} \;, \nonumber \\
P(\bar{\nu}_\mu \rightarrow \bar{\nu}_e)_{\rm LSND} & = & (2c_{12}s_{12})^2 c_{23}^2 \sin^2 (x _{21})_{\rm LSND} \simeq 0 \;, \nonumber \\ 
P(\bar{\nu}_e \rightarrow \bar{\nu}_e)_{\rm Chooz} & = & 1 - (2c_{12}s_{12})^2 \sin^2 (x _{21})_{\rm Chooz} \simeq 1\;, 
\end{eqnarray} 

\ni where $\Delta m^2_{21} \ll \Delta m^2_{32} \simeq \Delta m^2_{31}$ and

\begin{equation} 
x_{lk} = 1.27 \frac{\Delta m^2_{lk} L}{E} \;,\; \Delta m^2_{lk}  = m^2_l - m^2_k \;(k, l = 1,2,3)
\end{equation} 

\ni ($\Delta m^2_{lk}$, $L$ and $E$ are measured in eV$^2$, km and GeV, respectively). 

It is worthwhile to mention that the pseudo-Dirac mass spectrum can be derived from Eqs. (13) as the formal limit $ m_i = - \lim[{\stackrel{0}{m}}/(1- t^2_{ij})] = - m_j$ with $ t_{ij} \rightarrow 1$ and ${\stackrel{0}{m}} \rightarrow 0$ ({\it i.e.}, $c_{ij} \rightarrow 1/\sqrt{2} \leftarrow s_{ij}$). Then, it turns out that in our model also in the pseudo-Dirac case the nearly bimaximal oscillation formulae (22) hold. This is a consequence of $s_{13} = 0$ in $U^{(3)}$ and of the mass-squared degeneracy $ m^2_i = m^2_j \;(ij = 14, 25, 36) $.

Experimental estimations for solar $\nu_e$'s and atmospheric $\nu_\mu$'s  are $\theta_{12} \sim (32^\circ \;{\rm or}\; 38^\circ)$, $|\Delta m^2_{21}| \sim (5\times 10^{-5}$ or $7.9\times 10^{-8})\;{\rm eV}^2$ [7] and $\theta_{32} \sim 45^\circ$, $|\Delta m^2_{32}| \sim 2.5\times 10^{-3}\;{\rm eV}^2$ [8], respectively. For solar $\nu_e$'s they correspond to the MSW Large Mixing Angle solution or MSW LOW solution, respectively; the first is favored. The mixing angles give $c_{12} \sim (1.2/\sqrt{2} \;{\rm or}\; 1.1/\sqrt{2})$, $s_{12} \sim (0.75/\sqrt{2} \;{\rm or}\; 0.87/\sqrt{2})$ and $c_{23} \sim 1/\sqrt{2} \sim s_{23}$. The mass-squared differences are  hierarchical, $\Delta m^2_{21} \ll \Delta m^2_{32} \simeq \Delta m^2_{31}$, implying in the case of our Eqs. (13) the option of hierarchical mass spectrum $ m^2_1 < m^2_2 \ll m^2_3$ with $\Delta m^2_{32} \simeq  m^2_3$ and $\Delta m^2_{21}/ \Delta m^2_{32} \sim 2.0\times 10^{-2}$ or $3.2\times 10^{-5}$ (here, the ordering $ m^2_1 \leq m^2_2 \leq m^2_3$ is used). 

The rate of neutrinoless double $\beta$ decay (allowed only in the case of Majorana-type $\nu_{e L})$ is proportional to $m^2_{e e}$, where $m_{e e} \equiv |\sum_{i=1}^6 U^2_{e i} m_i|$ is reduced to $m_{e e} \simeq |\sum_{i=1}^3 U^{(3)\,2}_{e i} m_i| = c^2_{12}|m_1| + s^2_{12}|m_2| \sim (0.72 |m_1| + 0.28 |m_2| \;{\rm or}\; 0.62 |m_1| + 0.38 |m_2|)$ in the seesaw case of $c^2_{ij} \gg s^2_{ij}$ and to $m_{e e} = 0$ in the pseudo-Dirac case of $c^2_{ij} = 1/2 = s^2_{ij}$ and $m_i + m_j = 0\;(ij = 14,25,36)$. Since $|m_1| \leq |m_2|$, one obtains in the first case that $|m_1| \leq m_{e e} \leq |m_2|$. The suggested experimental upper limit for $m_{e e}$ is $m_{e e}\, \stackrel{<}{\sim}$ (0.35 --- 1) eV [9]. If the actual $m_{e e}$ lay near its upper limit, then the option of nearly degenerate spectrum $ m_1^2 \simeq m^2_2 \simeq m^2_3$ with  hierarchical mass-squared differences $\Delta m^2_{21} \ll \Delta m^2_{32} \simeq \Delta m^2_{31}$ would be favored.

\vspace{0.3cm}

\ni {\bf 3. Conclusions.} In this note, an explicit model of neutrino texture was presented, where in the overall $6\times 6$ mass matrix $M$ its lefthanded $3\times 3$ component $M^{(L)}$ is zero, its righthanded $3\times 3$ component $M^{(R)}$ is diagonal with equal entries and its Dirac $3\times 3$ component $M^{(D)}$ is given as a diagonal, potentially hierarchical structure, deformed by the popular, nearly bimaximal $3\times 3$ mixing matrix $U^{(3)}$. Before its deformation, such a Dirac structure may be similar in shape to the charged-lepton and quark $3\times 3$ mass matrices that, of course, are also of Dirac type. In this model, if $M^{(R)}$ dominates over $M^{(D)}$, the familiar seesaw mechanism works, leading effectively to the popular, nearly bimaximal oscillations of active neutrinos, governed by the mixing matrix $U^{(3)}$ involved in $M^{(D)}$.

In the presented model of neutrino texture, where 

$$ 
U = \stackrel{1}{U}\stackrel{0}{U}\;,\;\stackrel{0}{U}\!^\dagger \stackrel{0}{M} \stackrel{0}{U} = U^\dagger M U = {\rm diag}(m_1, m_2, m_3, m_4, m_5, m_6) \;,\; \stackrel{0}{M} = \stackrel{1}{U}\!^\dagger M {\stackrel{1}{U}}\;,
$$

\ni the following remarkable formulae hold in the seesaw approximation:

$$
U \simeq U_{\rm eff}\stackrel{1}{U}\;,\; \stackrel{1}{U}\!^\dagger M_{\rm eff} \stackrel{1}{U} \simeq U^\dagger M U = {\rm diag}(m_1, m_2, m_3, m_4, m_5, m_6) \;,\; M_{\rm eff} =U_{\rm eff}^\dagger M  U_{\rm eff}\;. 
$$

\ni Here, $M_{\rm eff}$ is the seesaw effective mass matrix approximately equal to the familiar block-diagonal form.

\vspace{0.3cm}

\ni {\bf 4. Outlook.} We find attractive the idea expressed in Eq. (7) that the Dirac component of neutrino overall mass matrix is similar in shape to the charged-lepton and quark mass matrices, before this component is deformed by the nearly bimaximal mixing. To proceed a bit further with this idea we will try to conjecture that this Dirac component has a shape analoguous to the following charged-lepton mass matrix [10]:

\begin{equation}  
{M}^{(e)} = \frac{1}{29} \left(\begin{array}{ccc} \mu^{(e)}\varepsilon^{(e)} & 0 & 0 \\ & & 
\\ 0 & 4\mu^{(e)}(80 + \varepsilon^{(e)})/9 & 0 \\ & & 
\\ 0 & 0 & 24\mu^{(e)} (624 + \varepsilon^{(e)})/25 \end{array}\right)  
\end {equation}  

\ni which predicts accurately the mass $m _\tau = M_{\tau \tau}^{(e)}$ from the experimental values of masses $m_e = M^{(e)}_{e e}$ and $m_\mu = M^{(e)}_{\mu \mu}$, when they are used as an input. In fact, we get then $m_\tau = 1776.80$ MeV [10] {\it versus} $m^{\rm exp}_\tau = 1777.03^{+0.30}_{-0.26}$ MeV [11] (and, in addition, $\mu^{(e)} =85.9924$ MeV and $\varepsilon^{(e)} = 0.172329$). For a theoretical background of this particular form of $M^{(e)}$ the interested reader may consult Ref. [12]. Let us emphasize that the figures in the mass matrix (24) are not fitted {\it ad usum Delphini}.

Thus, making use of Eqs. (15) and (9) as well as the neutrino analogue of Eq. (24) for ${\stackrel{0}{M}}\,\!^{(D)}$, we put

\begin{eqnarray}
{\stackrel{0}{m}} \frac{t_{14}}{1 - t^2_{14}} & = & \stackrel{0}{M}\!^{(D)}_{e e} = \frac{\mu^{(\nu)}}{29} \,\varepsilon^{(\nu)} \;, \nonumber \\
{\stackrel{0}{m}} \frac{t_{25}}{1 - t^2_{25}} & = & \stackrel{0}{M}\!^{(D)}_{\mu \mu} = \frac{\mu^{(\nu)}}{29} \frac{4(80+\varepsilon^{(\nu)})}{9} \;, \nonumber \\
{\stackrel{0}{m}} \frac{t_{36}}{1 - t^2_{36}} & = & \stackrel{0}{M}\!^{(D)}_{\tau \tau} = \frac{\mu^{(\nu)}}{29} \frac{24(624+\varepsilon^{(\nu)})}{25}\;.
\end{eqnarray}

\ni Hence, taking $\varepsilon^{(\nu)}\! = 0\!$ (already $\varepsilon^{(e)}$ is small) and anticipating that $\mu^{(\nu)}\!/{\stackrel{0}{m}} \ll \!1$, we calculate

\begin{equation}  
t_{14} = 0 \;,\;t_{25} = 1.23 \frac{\mu^{(\nu)}}{\stackrel{0}{m}} \;,\;t_{36} = 20.7 \frac{\mu^{(\nu)}}{\stackrel{0}{m}} \
\end {equation}  

\ni (note that the anticipation of $\mu^{(\nu)}/\stackrel{0}{m} \ll 1$ implies the choice of the seesaw case).  Then, from the first Eqs. (13)

\begin{equation}  
m^2_1 = 0 \;,\;m^2_2 = 2.26 \frac{\mu^{(\nu)\,4}}{\stackrel{0}{m}\!^2} \;,\;m^2_3 = 1.82\times 10^5 \frac{\mu^{(\nu)\,4}}{\stackrel{0}{m}\!^2} 
\end {equation}  

\ni and

\begin{equation}  
\Delta m^2_{21} = m^2_2 \;,\;\Delta m^2_{32}= m^2_3 - m^2_2 = 1.82 \times 10^5 \frac{\mu^{(\nu)\,4}}{\stackrel{0}{m}\!^2} \;,\;\Delta m^2_{21}/ \Delta m^2_{32} = 1.24\times 10^{-5}. 
\end {equation}

\ni Using in the second Eq. (28) the \Stk estimate $\Delta m^2_{32} \sim 2.5\times 10^{-3}\; {\rm eV}^2$ [8], we get

\begin{equation}  
\mu^{(\nu)\,4} \sim 1.4 \times 10^{-8} \stackrel{0}{m}\!^2\; {\rm eV}^2 \;,\; \mu^{(\nu)\,2} \sim 1.2 \times 10^{-4} \stackrel{0}{m} {\rm eV}\;.
\end {equation}  

If taking reasonably $\mu^{(\nu)} \leq \mu^{(e)} = 85.9924$ MeV, we obtain from Eq. (29) $ \stackrel{0}{m}\,\stackrel{<}{\sim} \, 6.3 \times 10^{10}$ GeV. Thus, in the case of maximalistic conjecture of $\mu^{(\nu)} = \mu^{(e)}$ (and consequently $\varepsilon^{(\nu)} = \varepsilon^{(e)} \simeq 0$) the mass scale is determined as $\stackrel{0}{m} \, \sim 6.3\times 10^{10}$ GeV, and then from Eqs. (26) $ t^2_{25} \sim 2.8\times 10^{-24}$ and $ t^2_{36} \sim 7.9\times 10^{-22}$. But, a dramatically smaller $\stackrel{0}{m}$ can also give $t^2_{ij} \ll 1$, {\it e.g.} for $\stackrel{0}{m}\, \sim 1$ eV we get $\mu^{(\nu)\,2} \sim 1.2 \times 10^{-4} {\rm eV}^2$ and thus, from Eqs. (26) $t^2_{25} \sim 1.8\times 10^{-4}$ and $t^2_{36} \sim 5.0\times 10^{-2}$. For such a low mass scale $\stackrel{0}{m} $ the three additional mass neutrinos $\nu_{j L}\;(j = 4,5,6)$ would be also light since $ m_j \simeq \,\stackrel{0}{m}\,$ for $t^2_{ij} \ll 1$, although $|m_i|/m_j = t^2_{ij} \ll 1$ [see Eqs. (13)]. This would not modify, however, the neutrino oscillations described in Eqs. (22) as long as $t^2_{ij} \ll 1$ and so, the seesaw works. Then, $\nu_{j L}\;(j =4,5,6)$ are approximately equal to $(\nu_{\alpha R})^c\;(\alpha = e, \mu, \tau)$ and decoupled from $\nu_{i L}\;(i = 1,2,3)$ which in turn are  nearly identical with $\nu_{\alpha L}\;(\alpha = e, \mu, \tau)$.

From the ratio $\Delta m^2_{21}/ \Delta m^2_{32} $ in Eq. (28) and the estimate $ \Delta m^2_{32} \sim 2.5\times 10^{-3}\;{\rm eV}^2$ we obtain the {\it prediction}

\begin{equation}  
m^2_2 = \Delta m^2_{21} \sim 3.1\times 10^{-8}\;{\rm eV}^2
\end {equation}

\ni which lies not so far from the experimental estimate $ \Delta m^2_{21} \sim 7.9\times 10^{-8}\;{\rm eV}^2$ based on the MSW LOW solar solution [7], whereas the favored experimental estimation based on the MSW Large Mixing Angle solar solution is much larger: $ \Delta m^2_{21} \sim 5\times 10^{-5}\;{\rm eV}^2$. So, if really true, the latter excludes dramatically the conjecture (25). Otherwise, this conjecture might be a significant step  forwards in our understanding of neutrino texture, in particular, of the question of fermion universality extended to neutrinos.

If the predictions $m^2_1 = 0$ and $m^2_2 \sim 3.1\times 10^{-8}\;{\rm eV}^2$ were true, then our previous estimate $m_{e e} \sim 0.62|m_1| + 0.38|m_2|$ of the effective mass of $\nu_e$ in the neutrinoless double $\beta$ decay would give $m_{e e} \sim 6.7\times 10^{-5}$ eV, much below the presently suggested experimental upper limit $m_{e e} \stackrel{<}{\sim}$ (0.35 --- 1) eV [9] (recall, however, that here $ U^{(3)}_{e 3} =0$). Thus, these predictions wouild imply the option of hierarchical neutrino spectrum $0 =m^2_1 < m^2_2 \ll m^2_3 \sim 2.5\times 10^{-3}\;{\rm eV}^2$ with the tiny $m^2_{e e} \sim 4.5\times 10^{-9}\;{\rm eV}^2$, much too small to allow for the detection of $0\nu \beta \beta$ decay in present experiments.

The Dirac component of the generic neutrino mass term (5),

\begin{eqnarray} 
- {\cal L}^{(D)}_{\rm mass} & = & \frac{1}{\sqrt{2}} \sum_{\alpha \beta} \left(\overline{\nu_{\alpha L})^c} \,,\, \overline{\nu_{\alpha R}}\right) \left( \begin{array}{cc} 0 & M^{(D)}_{\alpha \beta} \\ M^{(D)}_{\beta \alpha} & 0 \end{array} \right) \left( \begin{array}{c} \nu_{\beta L} \\ (\nu_{\beta R})^c \end{array} \right) + {\rm h.\,c.} \nonumber \\ & = & \sum_{\alpha \beta} \overline{\nu_{\alpha R}} M^{(D)}_{\beta \alpha} \nu_{\beta L} + {\rm h.\,c.} \;,
\end{eqnarray} 

\ni may arise from  the conventional doublet Higgs mechanism. In fact, writing in our model

\begin{equation}  
{M}^{(D)}_{\beta \alpha} = \sum_\gamma {U}^{(3)}_{\beta \gamma} \stackrel{0}{M}\!^{(D)}_{\gamma \alpha} = {U}^{(3)}_{\beta \alpha} \stackrel{0}{M} \!^{(D)}_{\alpha \alpha} = {U}^{(3)}_{\beta \alpha} Y^{(\nu)}_\alpha \langle \phi^0 \rangle
\end {equation}

\ni with $\stackrel{0}{M} \!^{(D)}_{\alpha \alpha} \;(\alpha = e,\mu,\tau)$ as given in Eqs. (25) and $\mu^{(\nu)} = \xi^{(\nu)} \langle \phi^0 \rangle $ {\it i.e.}, 

\begin{eqnarray} 
Y^{(\nu)}_e & = & \frac{\xi^{(\nu)}}{29}\, \varepsilon^{(\nu)} = 0 \;, \nonumber \\ Y^{(\nu)}_\mu & = & \frac{\xi^{(\nu)}}{29} \frac{4(80+\varepsilon^{(\nu)})}{9} = 1.23 \, \xi^{(\nu)} \;, \nonumber \\ 
Y^{(\nu)}_\tau & = & \frac{\xi^{(\nu)}}{29} \frac{24(624+\varepsilon^{(\nu)})}{25} = 20.7 \, \xi^{(\nu)} 
\end{eqnarray} 

\ni (for $\varepsilon^{(\nu)} = 0$), we obtain 

\begin{equation}  
- {\cal L}^{(D)}_{\rm mass} = \sum_{\alpha \beta} f^{(\nu)}_{\beta \alpha}\, \langle \phi^0 \rangle\, \overline{\nu_{\alpha R}}\, \nu_{\beta L} + {\rm h.\,c.} \;,
\end {equation}

\ni where

\begin{equation}  
f^{(\nu)}_{\beta \alpha} = U^{(3)}_{\beta \alpha} Y^{(\nu)}_{\alpha}  \;.
\end {equation}

\ni Here, $\xi^{(\nu)} = \mu^{(\nu)}/\langle \phi^0 \rangle $ with $\langle \phi^0 \rangle = 246.22$ GeV [13]. If  $\mu^{(\nu)} \leq \mu^{(e)} = 85.9924$ MeV, then $\xi^{(\nu)} \leq 3.4925\times 10^{-4}$. The Dirac component (34) of $6\times 6$ neutrino mass term arises from the following doublet Higgs-neutrino coupling term:

\begin{equation}  
- {\cal L}^{(\nu)}_{\phi} = \sum_{\alpha \beta} f^{(\nu)}_{\beta \alpha}\, \overline{\nu_{\alpha R}}\, (\nu_{\beta L}\phi^0 - e^-_{\beta L}\phi^+) + {\rm h.\,c.} \;,
\end {equation}

\ni when ${\cal L}^{(\nu)}_{\phi} \rightarrow {\cal L}^{(\nu)}_{\langle{\phi}\rangle}$ with $\phi^0 \rightarrow \langle \phi^0 \rangle $ and $\phi^+ \rightarrow \langle \phi^+ \rangle = 0$. The more familiar doublet Higgs-charged lepton coupling term is

\begin{equation}  
- {\cal L}^{(e)}_{\phi} = \sum_{\alpha \beta} f^{(e)}_{\alpha \beta} (\overline{\nu_{\alpha L}}\, \phi^+ + \overline{e^-_{\alpha L}}\, \phi^0)\, e^-_{\beta R} + {\rm h.\,c.} 
\end {equation}

\ni with $f^{(e)}_{\alpha \beta} = Y^{(e)}_\alpha \delta_{\alpha \beta}$ (in the flavor representation used here the charged-lepton mass matrix is diagonal). Here, $Y^{(e)}_\alpha $ is given in Eqs. (33), when $\xi^{(\nu)}$ and $\varepsilon^{(\nu)}$ are replaced there by $\xi^{(e)} = \mu^{(e)}/\langle \phi^0 \rangle = 3.4925\times 10^{-4}$ and $\varepsilon^{(e)} = 0.172329$, respectively.The arising charged-lepton masses are $ m_{e_\alpha} = Y^{(e)}_\alpha \langle \phi^0 \rangle = M_{\alpha \alpha}$ with $\xi^{(e)} \langle \phi^0 \rangle = \mu^{(e)} = 85.9924$ MeV ($\alpha = e,\mu,\tau$, $m_{e_\alpha} = m_{e},m_\mu,m_\tau$), what gives $m_e = 0.510999$ MeV, $m_\mu = 105.658$ MeV, $m_\tau = 1776.80$ MeV (the experimental values of $m_e$ and $m_\mu $ were inputs to evaluate $\mu^{(e)}$ and $\varepsilon^{(e)}$ and predict $m_\tau $).

Finally, we would like to mention that if in our model there were $\!M^{(L)} = \,{\stackrel{0}{m}}1^{(3)}$,    $M^{(R)} = 0$ and $M^{(D)} = -{\stackrel{0}{m}}\, U^{(3)} \frac{1}{2}$ diag$ (\tan 2\theta_{14}, \tan 2\theta_{25}, \tan 2\theta_{36})$ [6], leading to the same $U$ as in Eqs. (11) and (12) but to the interchanged $m_i \leftrightarrow m_j$ in Eqs. (13), then the predicted $\Delta m^2_{21}$ would be of the order of $10^{-5}$ eV$^2$, not very far from its favored experimental estimate $5\times 10^{-5}$ eV$^2$ based on the MSW Large Mixing Angle solar solution (now, $\mu^{(\nu)\,2} \sim 2.9\times 10^{-6}$ eV$^2$).

\vfill\eject

~~~~
\vspace{0.5cm}

{\centerline{\bf References}}

\vspace{0.45cm}

{\everypar={\hangindent=0.6truecm}
\parindent=0pt\frenchspacing

{\everypar={\hangindent=0.6truecm}
\parindent=0pt\frenchspacing

~[1]~{\it Cf. e.g.} Z. Xing, {\it Phys. Rev.} {\bf D 61}, 057301 (2000); and references therein.

\vspace{0.2cm}

~[2]~Z. Maki, M. Nakagawa and S. Sakata, {\it Prog. Theor. Phys.} {\bf 28}, 870 (1962).

\vspace{0.2cm}

~[3]~For a recent review {\it cf.} M.C. Gonzalez-Garcia and Y.~Nir, {\tt hep--ph/0202056}; and references therein.

\vspace{0.2cm}

~[4]~~M. Gell-Mann, P. Ramond and R.~Slansky, in {\it Supergravity}, edited by F.~van Nieuwenhuizen and D.~Freedman, North Holland, 1979; T.~Yanagida, Proc. of the {\it Workshop on Unified Theory and the Baryon Number in the Universe}, KEK, Japan, 1979; R.N.~Mohapatra and G.~Senjanovi\'{c}, {\it Phys. Rev. Lett.} {\bf 44}, 912 (1980).

\vspace{0.2cm}

~[5]~{\it Cf. e.g.} W. Kr\'{o}likowski, {\it Acta Phys. Pol.} {\bf B 30}, 663 (2000); and references therein. 

\vspace{0.2cm}

~[6]~For a previous version {\it cf.} W. Kr\'{o}likowski, {\it Acta Phys. Pol.} {\bf B 33}, 641 (2002). 

\vspace{0.2cm}

~[7]~V. Barger, D. Marfatia, K. Whisnant and B.P.~Wood, {\tt hep--ph/0204253}; J.N.~Bahcall, M.C.~Gonzalez--Garcia and C. Pe\~{n}a--Garay, {\tt hep--ph/0204314v2}; and references therein.

\vspace{0.2cm}

~[8]~S. Fukuda {\it et al.}, {\it Phys. Rev. Lett.} {\bf 85}, 3999 (2000).

\vspace{0.2cm}

~[9]~{\it Cf. e.g.} M. Frigerio and A.Yu. Smirnov, {\tt hep--ph/0202247}; and references therein.

\vspace{0.2cm}

[10]~W. Kr\'{o}likowski, {\it Acta Phys. Pol.} {\bf B 27}, 2121 (1996); and references therein.

\vspace{0.2cm}

[11]~The Particle Data Group, {\it Eur. Phys. J.} {\bf C 15}, 1 (2000).

\vspace{0.2cm}

[12]~W. Kr\'{o}likowski, Appendix in {\it Acta Phys. Pol.} {\bf B 32}, 2961 (2001); Appendix B in 
{\tt hep--ph/0201004v2}; {\tt hep--ph/0203107}; and references therein.

\vspace{0.2cm}

[13]~{\it Cf. e.g.} S. Weinberg, {\it The Quantum Theory of Fields}, Vol. II, Cambridge University Press, 1996.

\vfill\eject

\end{document}